\documentclass{ws-procs9x6}
\newcommand{\ba}{\begin{eqnarray}}
\newcommand{\ea}{\end{eqnarray}}
\newcommand{\bmath}{\begin{mathletters}}
\newcommand{\emath}{\end{mathletters}}
\newcommand{\ban}{\begin{eqnarray*}}
\newcommand{\ean}{\end{eqnarray*}}

\newcommand{\bsub}{\begin{subequations}}
\newcommand{\esub}{\end{subequations}}

\begin{document}

\title{Quantum Phase Transitions in Odd-Mass Nuclei}

\author{A. Leviatan}

\address{Racah Institute of Physics, The Hebrew University,\\ 
Jerusalem 91904, Israel\\
E-mail: ami@phys.huji.ac.il}

\author{D. Petrellis}

\address{Institute of Nuclear Physics, N.C.S.R. ``Demokritos'',\\ 
GR-15310 Aghia Paraskevi, Attiki, Greece\\
E-mail: petrellis@inp.demokritos.gr}

\author{F. Iachello}

\address{Center for Theoretical Physics, Sloane Physics Laboratory, 
Yale University,\\ 
New Haven, Connecticut~06520-8120, USA\\
E-mail: francesco.iachello@yale.edu}

\begin{abstract}
Quantum shape-phase transitions in odd-even nuclei are investigated 
in the framework of the interacting boson-fermion model. Classical 
and quantum analysis show that the presence of the odd fermion 
strongly influences the location and nature of the phase transition, 
especially near the critical point. Experimental evidence for 
the occurrence of spherical to axially-deformed transitions in 
odd-proton nuclei Pm, Eu and Tb (Z=61, 63, 65) is presented.
\end{abstract}

\bodymatter

\vspace{1cm} 

Quantum phase transitions (QPTs) are qualitative changes in the ground 
state properties of a physical system induced by a variation of 
parameters in the quantum Hamiltonian. 
These structural modifications have 
found a variety of applications in diverse areas of physics~\cite{carr}. 
One of these applications is to atomic nuclei, where QPTs 
in even-even nuclei have been
extensively investigated (for a review, see~\cite{iac3,casten11,iac11}) 
within the framework of the Interacting Boson Model (IBM)~\cite{iac1}, 
a model of nuclei in terms of correlated pairs of valence nucleons 
with $L=0,2$ treated as ($s, d$) bosons~\cite{iac1}. 
In the present contribution we extend these studies to odd-even nuclei 
making use of the Interacting Boson Fermion Model (IBFM)~\cite{iac2}, 
incorporating an additional unpaired fermion with angular momentum $j$. 
The reported results portray 
general properties of QPTs in mixed Bose-Fermi systems~\cite{PLI11} 
and are illustrated for $j=11/2$. 

We consider the Hamiltonian of a system of $N$ monopole ($s^{\dag}$) and 
quadrupole ($d^{\dag}_{\mu}$) bosons and a single-$j$ fermion 
($a^{\dag}_{j,m}$)
\ba
\hat{H} = \hat{H}_{B}+ \hat{H}_{F}+ \hat{V}_{BF}~,
\label{Eq:hBF}
\ea
with
\bsub
\ba
\hat{H}_{B} &=& 
\varepsilon _{0}\left[ \left( 1-\xi \right) \hat{n}_{d}-\frac{\xi }{
4N}\hat{Q}^{\,\chi }\cdot \hat{Q}^{\,\chi }\right] ~,\\  
\hat{H}_{F} &=&\varepsilon _{j}\,\hat{n}_{j} ~,\\
\hat{V}_{BF}&=& 
\Gamma\, \hat{Q}^{\chi }\cdot ( a_{j}^{\dag }\, \tilde{a}_{j} )^{(2)}
+ \Lambda\, \sqrt{2j+1}:[ ( d^{\dag }\, \tilde{a}_{j})^{(j)}
( \tilde{d}\, a_{j}^{\dag })^{(j)}]^{(0)}:~.
\qquad
\label{Eq:hBhFvBF}
\ea
\esub
Here $\hat{n}_{d}= d^{\dag }\cdot \tilde{d}$ and $\hat{n}_j$
are the $d$-boson and fermion number operators respectively,
and $\hat{Q}^{\,\chi }=( d^{\dag }s+s^{\dag } \tilde{d})^{(2)}
+\chi ( d^{\dag } \tilde{d})^{(2)}$. 
The parameter $\varepsilon _{0}$ is the scale of the boson energy, 
$\varepsilon _{j}$ 
is the energy of the single fermion, $\Gamma$ and $\Lambda$ are, 
respectively, the strengths of the quadrupole and exchange 
Bose-Fermi interactions. 
QPTs of the purely bosonic part of the Hamiltonian $\hat{H}_{B}$ have been 
extensively investigated~\cite{iac3,casten11,iac11}. 
There are two control parameters $\xi$ and $\chi $. 
For fixed $\chi $, as one varies $\xi $, $0\leq \xi \leq 1$,
the bosonic system undergoes a QPT. The phase transition is first order for 
$\chi \neq 0$ and becomes second order at $\chi =0$. No phase transition
occurs as a function of $\chi $. In this contribution, we take 
$\chi =-\frac{\sqrt{7}}{2}$, in which case the spherical phase has 
U(5) symmetry ($\xi =0$) and the axially-deformed phase has 
SU(3) symmetry ($\xi =1$)~\cite{iac1}. 
The critical point, separating the two phases 
occurs at $\xi_{c}\cong 1/2$. 

A complete study of the properties of quantum phase transitions 
necessitates both a classical and a quantal analysis. 
A classical analysis amounts to constructing the combined Bose-Fermi
potential energy surface (Landau potential) and minimizing it with respect
to the classical variables. To this end, we introduce a 
boson condensate~\cite{iac1}
\ba
\left\vert N;\beta ,\gamma \right\rangle = 
(N!)^{-1/2}\left[
b_{c}^{\dag }\left( \beta ,\gamma \right) \right] ^{N}
\left\vert 0\right\rangle~,
\label{Eq:cond}
\ea
$b_{c}^{\dag} = (1+\beta ^{2})^{-1/2}[ \beta\cos \gamma d_{0}^{\dag } 
+ \beta \sin \gamma (d_{2}^{\dag }+d_{-2}^{\dag })/\sqrt{2} +s^{\dag }]$, 
in terms of the classical variables $\beta ,\gamma$. 
By integrating out the boson degrees of freedom, i.e., by taking the
expectation value of $\hat{H}$~(\ref{Eq:hBF}) 
in the boson condensate, 
${\cal H}(N;\beta ,\gamma ) = 
\langle N;\beta ,\gamma \vert
\hat{H}\vert N;\beta ,\gamma \rangle $, 
one obtains the fermion Hamiltonian 
\ba
{\cal H}(N;\beta ,\gamma ) 
= E_{B}(N;\beta ,\gamma ) + \varepsilon _{j}\,\hat{n}_{j}
+ \sum_{m_{1},m_{2}} g_{m_{1},m_{2}}(N;\beta ,\gamma )  
a_{j,m_{1}}^{\dag }a_{j,m_{2}}~,
\qquad
\label{Eq:Hdef}
\ea
where $E_B = \langle N;\beta ,\gamma \vert
\hat{H}_{B}\vert N;\beta ,\gamma \rangle $. 
The matrix $g_{m_{1},m_{2}}(N;\beta ,\gamma )$ is a real, 
symmetric matrix, which depends on the Bose-Fermi couplings, 
$\Gamma$  and $\Lambda$. 
Its numerical diagonalization yields the single particle eigenvalues 
$e_{i}(\beta ,\gamma)$ and eigenfunctions 
$\psi _{i}(\beta ,\gamma)$, $ i=1,2,...,j+\frac{1}{2}$.
These are the single-particle levels in the deformed ($\beta ,\gamma $) 
field generated by the bosons. 
For $\gamma = 0^{\circ}$ 
the eigenvalues are given in explicit analytic form~\cite{lev88}
\ba
\lambda_{K}(\beta ) &=& -N\Gamma \left\{ \left( \frac{\beta }{1+\beta^{2}}
\right) \sqrt{5}\left (\, 2-\beta \chi \sqrt{\frac{2}{7}}\,\right )
P_{j}[3K^{2}-j(j+1)]\right\}   
\nonumber \\
&&-N\Lambda \left\{ \left( \frac{\beta ^{2}}{1+\beta ^{2}}\right)
(2j+1)P_{j}^{2}[3K^{2}-j(j+1)]^{2}\right\}~,
\qquad
\ea
where $P_{j}=\left[ (2j-1)j(2j+1)(j+1)(2j+3)\right] ^{-1/2}$. 
The eigenvalues can be 
labelled by the angular momentum projection on
the intrinsic axis $\hat{3}$, $K_{3}\equiv K=\frac{1}{2},\frac{3}{2},...,j$
and they are doubly degenerate. 
Similarly, the eigenvalues for  $\gamma=60^{\circ}$  (oblate axial symmetry),
are given by $\omega_{K_2}(\beta) = \lambda_{K_3\to K_2}(-\beta)$, 
where $K_2$ is the corresponding projection on the axis $\hat{2}$. 
Once the eigenvalues have been obtained, one can calculate the total energy
functional (Landau potential for the combined Bose-Fermi system)
\ba
E_{i}(N;\beta ,\gamma ;\xi ,\chi ;\Gamma ,\Lambda )
= E_{B}(N;\beta ,\gamma;\xi ,\chi )+\varepsilon _{j} + 
e_{i}(N;\beta ,\gamma ;\chi ;\Gamma ,\Lambda)~.
\label{Eq:Ei}
\ea
Minimization of $E_{i}$ with respect to $\beta$ and $\gamma $ gives the
equilibrium values $\beta_{e},\gamma_{e}$ 
(the classical order parameters) for each state. 
In general, $E_B\propto N^2$ and $e_i\propto N$, however, as shown below,
contrary to naive expectations, the odd fermion greatly influences the 
phase transition.

To facilitate the study of QPTs in the IBFM we set~\cite{alonso} 
$\Gamma = -2\varepsilon_{0}\frac{\xi }{4N}\,Q_{jj}$
with $Q_{jj}= \langle j\vert\vert Y^{(2)}\vert\vert j\rangle$ 
and examine the change in ($\beta_{e},\gamma_{e}$) upon variation 
of the control parameter $\xi$. 
\begin{figure}[t]
\begin{center}
\psfig{file=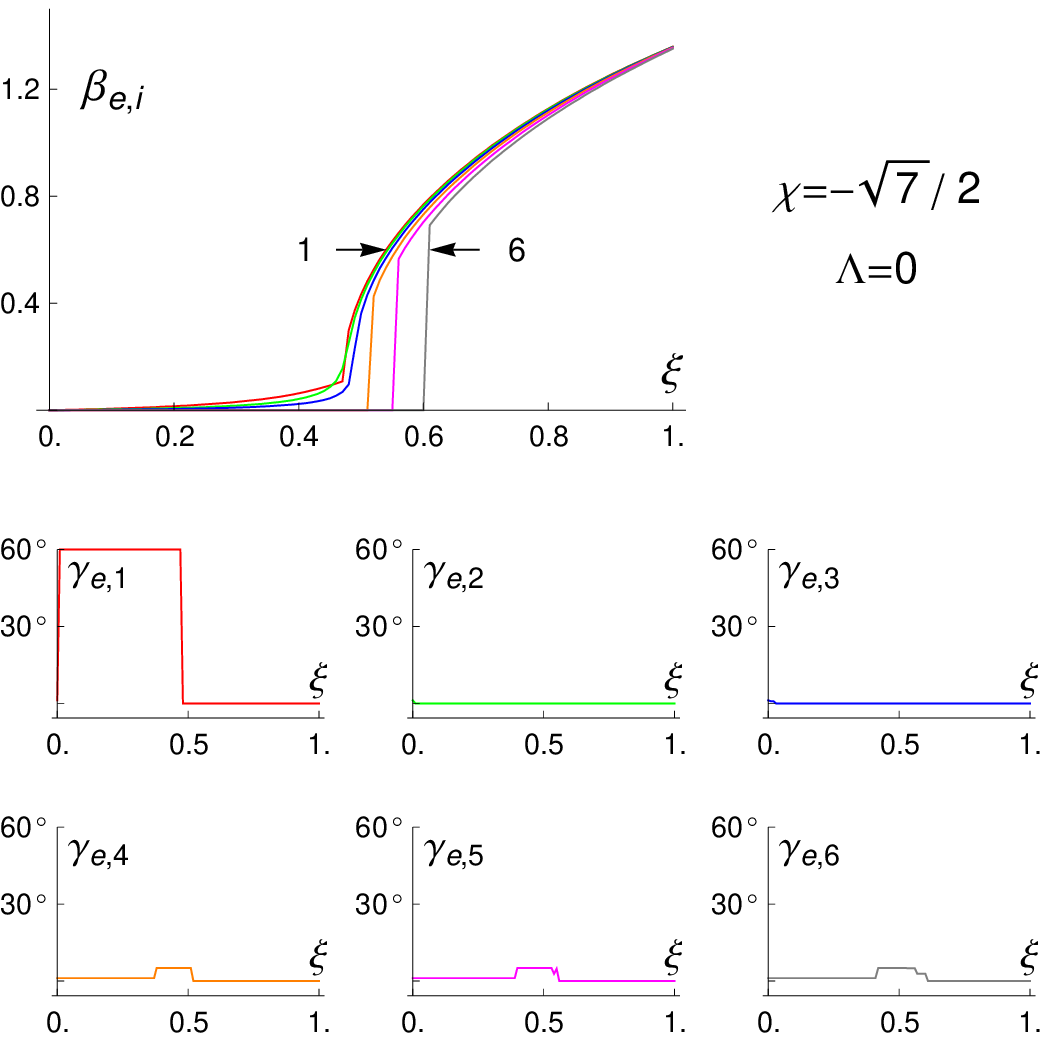,width=2.1in}
\hspace{0.4cm}
\psfig{file=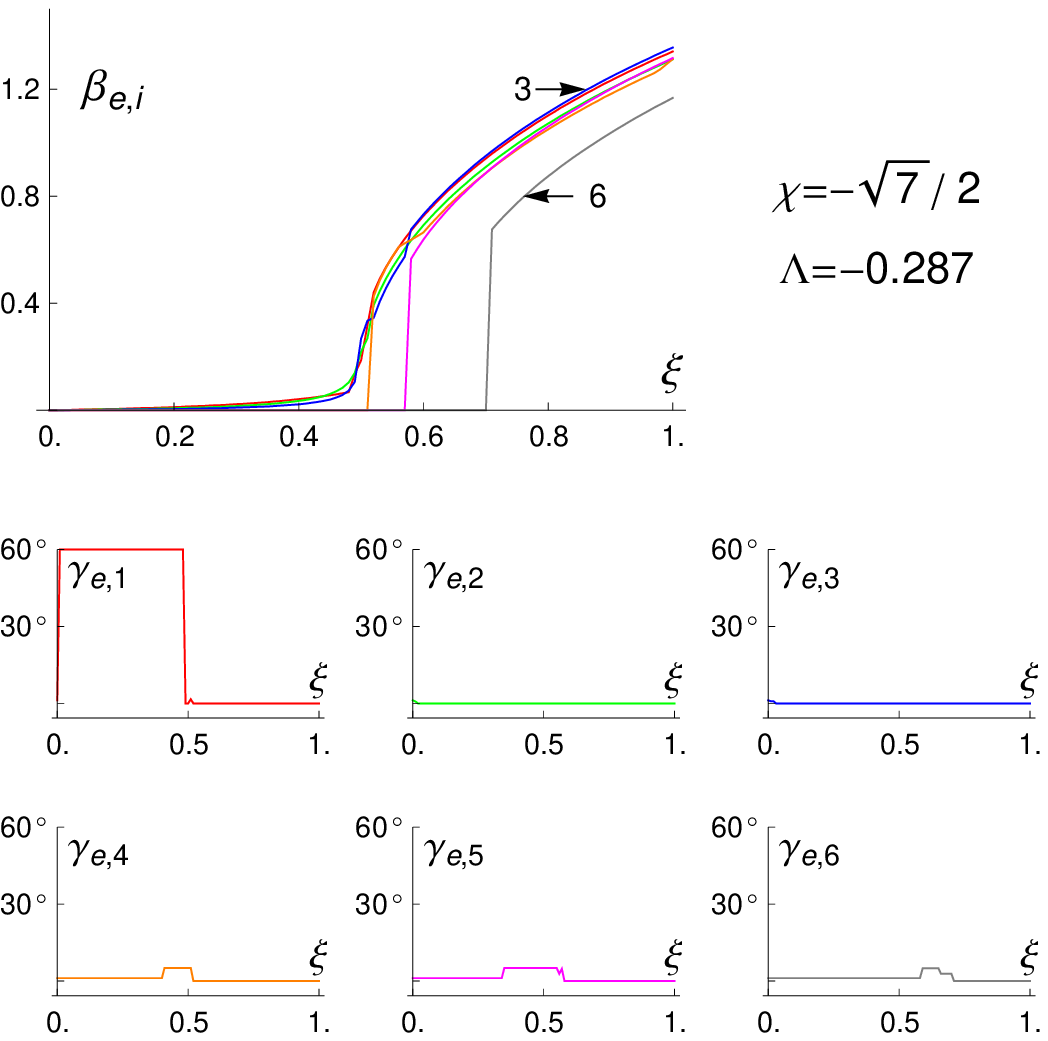,width=2.1in}
\end{center}
\caption{Equilibrium values, $\beta _{e,i}$ (top part) and $\gamma _{e,i}$
(bottom part) as a function of the control parameter $\xi $ in the
U(5)-SU(3) transition ($\chi =-\frac{\sqrt{7}}{2}$), 
and $N=10$. Left: $\Lambda =0$. Right: $\Lambda = -0.287$.  
States are labelled by the index $i=1,...,6$.}
\label{fig1}
\end{figure}
For pure quadrupole coupling ($\Lambda=0$), one can see from the left 
panel of Fig.~\ref{fig1} that the phase transition is washed out for 
states~1,2,3 and enhanced for states~4,5,6. The critical point is 
approximately at the same location as for the purely bosonic case 
($\xi_{B}^{c}\sim$0.51) for state~4 ($\xi _{4}^{c}\sim $0.50), but it 
is moved to larger values for state~5 ($\xi _{5}^{c}\sim $0.53) and state~6 
($\xi_{6}^{c}\sim $0.58). The values of $\gamma _{e}$ below and around the
critical point are no longer zero. After the critical point, all states
become those of a single particle in an axially deformed field 
($\gamma_{e}=0^{\circ }$) with equal deformation 
$\beta _{e,i}=\beta _{e,B}$ ($\beta _{e,B}$ being the equilibrium 
deformation of $E_B$). States in this region can be labelled by 
$K=\frac{1}{2},\frac{3}{2},...\frac{11}{2}$, 
corresponding to states~$i=1,...,6$. 
The equilibrium values when $\Lambda =-0.287$ 
are shown in the right panel of Fig.~\ref{fig1}. 
We see that the modifications induced by the presence of a fermion to the
phase transition are more dramatic when the exchange interaction is added, 
especially in the location of the critical point $\xi _{i}^{c}$. 
The state~6 moves considerably to the right of $\xi _{B}^{c}$. 
In summary, QPTs in the
presence of an odd fermion have a different behavior as a function of the
control parameter $\xi $ than in the purely bosonic system. The odd fermion
acts as a catalyst for some states and as a retardative for others.
\begin{figure}[t]
\begin{minipage}{0.49\linewidth}
\psfig{file=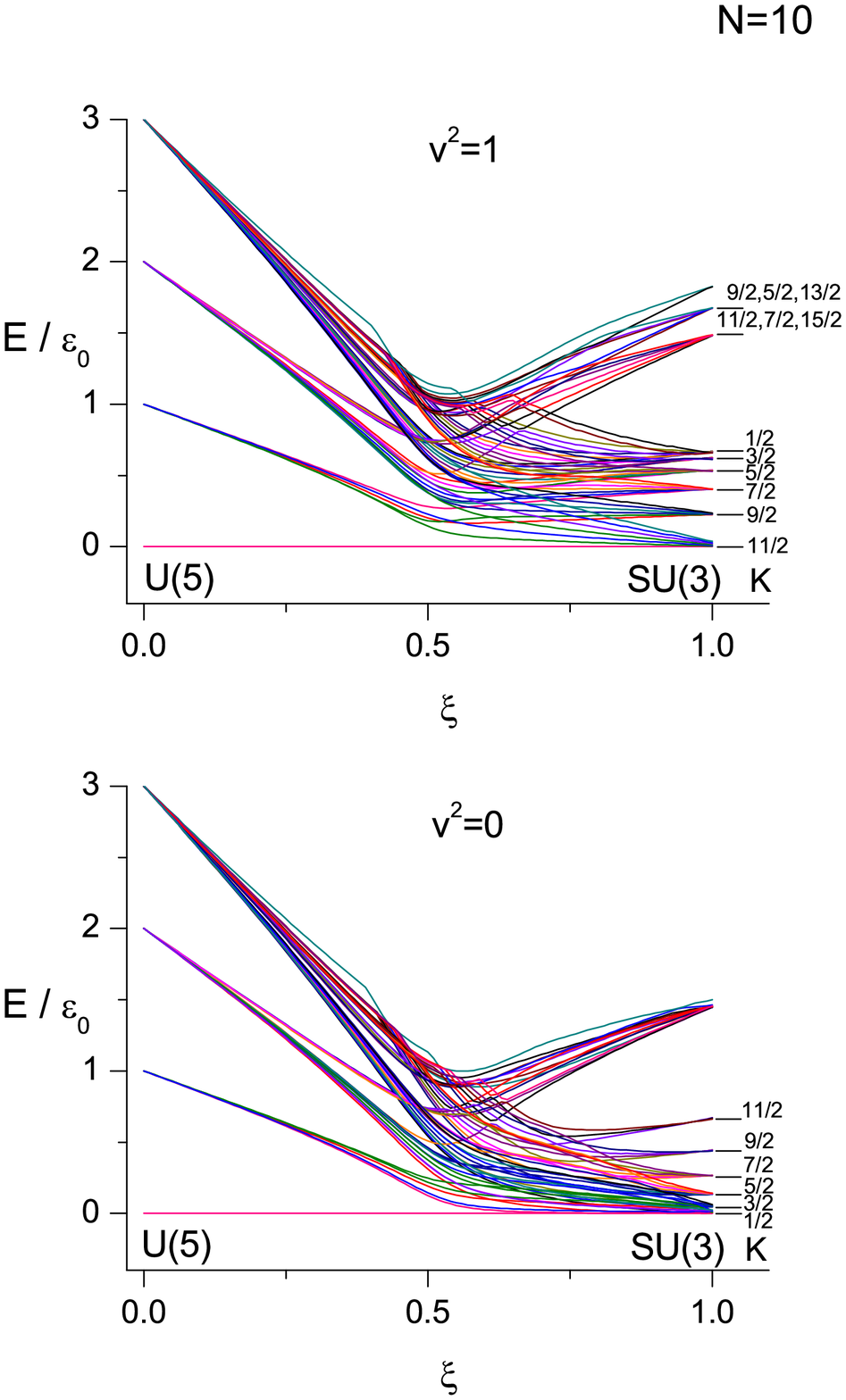,width=2in}
\caption{
Correlation diagram for a $j=11/2$ particle coupled to a system of 
($s,d$) bosons undergoing a U(5)-SU(3) phase transition. 
Top part $v^{2}=1$,
bottom part $v^{2}=0$. The interaction is purely quadrupole.}
\label{fig2}
\end{minipage}
\hspace{0.06cm}
\begin{minipage}{0.49\linewidth}
\psfig{file=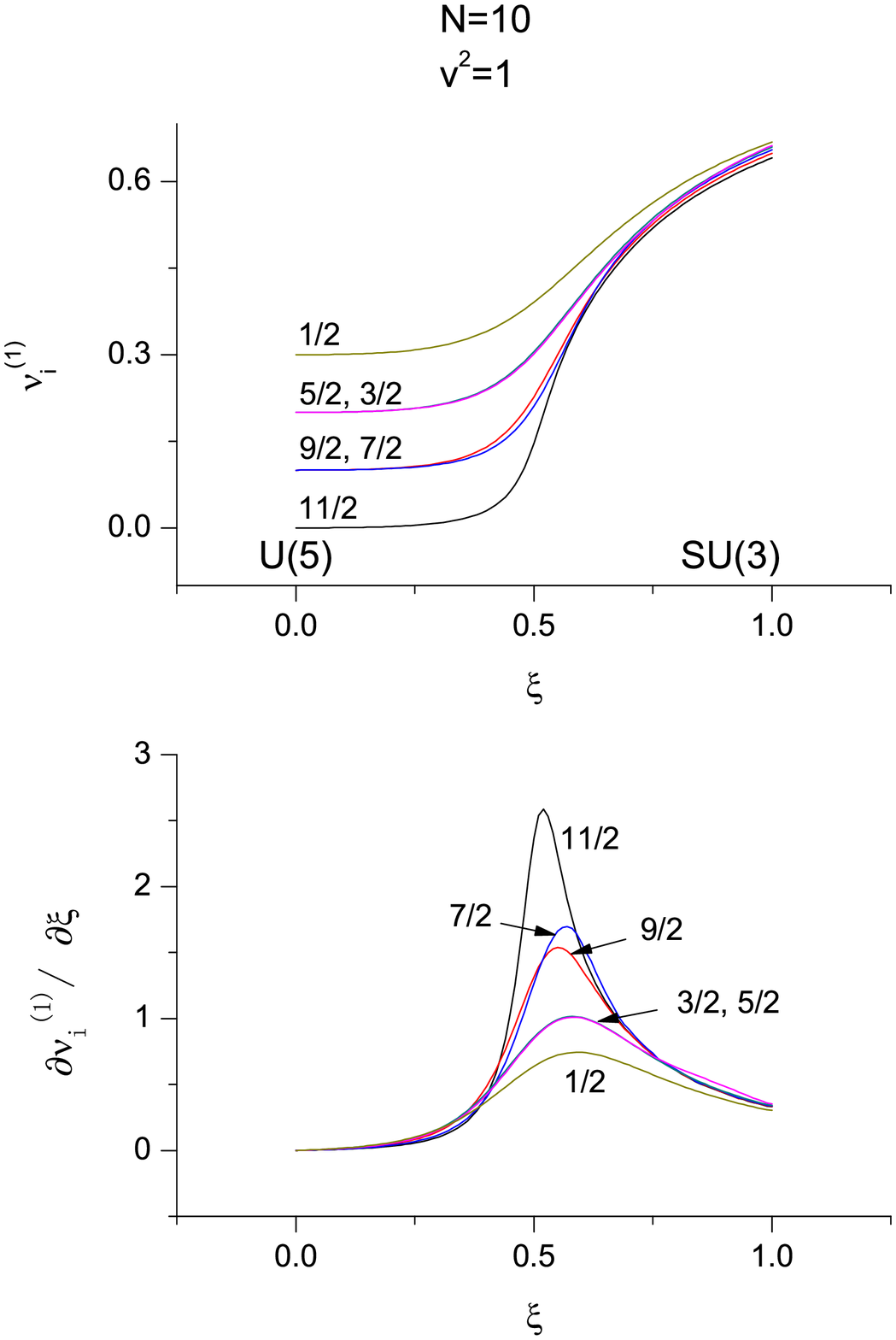,width=2.33in}
\caption{
The quantal order parameters, $\nu_{i}^{(1)}$, 
Eq.~(\ref{Eq:orderparam1}) (top part), 
and its derivative, 
(bottom part), for yrast states with $J=\frac{11}{2},..,\frac{1}{2}$, 
as a function of the control parameter, $\xi $. 
Here $N=10$ and $v^2=1$. 
\label{fig3}} 
\end{minipage}
\end{figure}

The quantal analysis of QPTs is done by diagonalizing numerically 
the Hamiltonian~(\ref{Eq:hBF}) for finite $N$. 
In the semi-microscopic version of the IBFM, 
the Bose-Fermi interactions are given in terms of the 
BCS occupation probabilities, $u_{j}$ and $v_{j}$, with $
u_{j}^{2}+v_{j}^{2}=1$. Specifically, 
$\Gamma =\Gamma _{s}\, 2\left( u_{j}^{2}-v_{j}^{2}\right) Q_{jj}$, 
$\Lambda = -\Lambda _{s}\, 8\sqrt{5}u_{j}^{2}v_{j}^{2}Q_{jj}^{2}/(2j+1)$ 
and we set $\Gamma _{s} = \frac{\xi}{4N}\,\varepsilon _{0}$. 
The correlation diagrams for pure quadrupole coupling ($\Lambda=0$) 
are presented in Fig.~\ref{fig2}. 
They describe how the energy levels evolve from the U(5) 
phase to the SU(3) phase. They also display particle-hole conjugation, 
that is the transformation ($u\leftrightarrow v$). 
As quantal order parameters we consider here the
expectation value of $\hat{n}_{d}$ in the states $i=1,...,6$, 
\ba
\nu _{i}^{(1)} = \frac{\left\langle \Psi _{i}
\left\vert \hat{n}_{d}\right\vert\Psi _{i}\right\rangle }{N} ~.
\label{Eq:orderparam1}
\ea
This is shown in Fig.~\ref{fig3} top part.
Since the order parameters $\nu _{i}^{(1)}$ are related to the square of the
classical order parameters $\beta _{e,i}$, this figure is related to 
Fig.~\ref{fig1} to which it corresponds in the limit $N\rightarrow \infty $. 
The derivative of $\nu _{1}^{(1)}$ in the ground state, 
$\frac{\partial \nu _{i}^{(1)}}{\partial \xi }$ is also shown in 
Fig.~\ref{fig3} bottom part. (This quantity
diverges when $N\rightarrow \infty $). From this figure one sees clearly
that the transition is made sharper by the presence of the fermion for some
states, $11/2,9/2,7/2$, while is made smoother for others, $5/2,3/2,1/2$, a
result already seen in the classical analysis.
\begin{figure}[t]
\begin{minipage}{0.5\linewidth}
\psfig{file=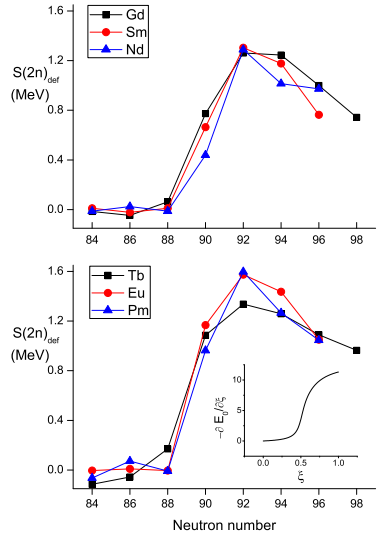,width=2in}
\end{minipage}
\begin{minipage}{0.45\linewidth}
\caption{
\label{fig4}
The contribution of deformation to the two-neutron separation
energies, $S(2n)_{\mathrm{def}}$ for even-even 
$_{60}$Nd-$_{62}$Sm-$_{64}$Gd nuclei (top) 
and odd-even $_{61}$Pm-$_{63}$Eu-$_{65}$Tb nuclei (bottom), plotted 
as a function of neutron number. The contribution is enhanced in odd-even
nuclei by approximately $300$ keV (at neutron number 92). Also the rise
between neutron numbers 88 and 90 is sharper in odd-even nuclei than in
even-even nuclei. In the limit $N\rightarrow \infty $ (no finite size
scaling) the quantity $S(2n)_{\mathrm{def}}$ should be zero before the 
critical value and finite and large after that. The expected behaviour of 
$-\frac{\partial E_0}{\partial \xi}$ for the U(5)-SU(3) transition and 
$N=10$ is shown in the inset.}
\end{minipage}
\end{figure}

One of the experimental signatures of QPT in nuclei 
is the two-neutron separation energies, 
$S_{2n} = -[E_{0}(N + 1) - E_{0}(N)]$, 
which can be related to the derivative of the
ground state energy, $E_0$, with respect to the control parameter, 
$\frac{\partial E_{0}}{\partial \xi }$. $S_{2n}$ 
can be written as a smooth contribution linear in the boson number $N$, 
plus the contribution of the deformation~\cite{iac1}
\ba
S_{2n} = -A_{2n}-B_{2n}N+S(2n)_{\mathrm{def}}~.
\label{Eq:S2n}
\ea
In order to emphasize the occurrence of the phase transition it is
convenient to plot the deformation contribution only, obtained from the 
data by subtracting the linear dependence, as a function of $N$. 
In previous studies of the purely bosonic part it has been shown that 
$N$ is approximately proportional to the control parameter 
$\xi$~\cite{iac3}. The experimental values of $S(2n)_{\mathrm{def}}$ 
are shown in the top part of Fig.~\ref{fig4} for even-even nuclei 
(purely bosonic) and in the bottom part for odd-even nuclei 
(bosonic plus one fermion). 
They are obtained from the empirical data with 
$A_{2n}=-14.61,\,-15.82,\, -16.997$ MeV for Nd-Sm-Gd, respectively, 
and $B_{2n}=0.657$ MeV, and with $A_{2n}=-15.185,\,-16.37,\,-17.672$ MeV 
for Pm-Eu-Tb, and $B_{2n}=0.670$ MeV. Precursors of the phase transition are
visible in all six nuclei between neutron numbers 88 and 90 in both, 
and, most importantly, appears to be enhanced in odd-even nuclei relative 
to the even-even case. 

This work was supported in part by the U.S.-Israel Binational Science 
Foundation and in part by DOE\ Grant No. DE-FG-02-91ER40608.

\end{document}